\documentclass[aps,prd,twocolumn,showpacs,amsmath,amssymb]{revtex4-1}
\usepackage{amsmath}
\usepackage{graphicx}
\usepackage{subfigure}
\usepackage{epstopdf}
\usepackage{color}
\usepackage{multirow}
\usepackage{setspace}
\usepackage{overpic}
\usepackage{amssymb}
\usepackage[bookmarksnumbered, pdfstartview=FitH,colorlinks,urlcolor=blue, citecolor=blue,linkcolor=blue] {hyperref}
\usepackage{lineno}
\usepackage{bm}
\usepackage{rotating}
\usepackage[utf8]{inputenc}

\usepackage{natbib}
\usepackage{mciteplus}

\let\oldequation\equation
\let\oldendequation\endequation

\renewenvironment{equation}
  {\linenomathNonumbers\oldequation}
  {\oldendequation\endlinenomath}

\begin{document}

\title{\bf \boldmath
	Observation of $D\to \bar{K}_{1}(1270)\mu^+\nu_\mu$ and test of lepton flavor universality with $D\to \bar{K}_1(1270) \ell^{+} \nu_{\ell}$
}

\author{M.~Ablikim$^{1}$, M.~N.~Achasov$^{4,c}$, P.~Adlarson$^{76}$, O.~Afedulidis$^{3}$, X.~C.~Ai$^{81}$, R.~Aliberti$^{35}$, A.~Amoroso$^{75A,75C}$, Q.~An$^{72,58,a}$, Y.~Bai$^{57}$, O.~Bakina$^{36}$
% ...
\\
\vspace{0.2cm}
(BESIII Collaboration)\\
\vspace{0.2cm} {\it
$^{1}$ Institute of High Energy Physics, Beijing 100049, People's Republic of China\\
% ...
$^{3}$ Bochum  Ruhr-University, D-44780 Bochum, Germany\\
$^{4}$ Budker Institute of Nuclear Physics SB RAS (BINP), Novosibirsk 630090, Russia\\
% ...
$^{35}$ Johannes Gutenberg University of Mainz, Johann-Joachim-Becher-Weg 45, D-55099 Mainz, Germany\\
$^{36}$ Joint Institute for Nuclear Research, 141980 Dubna, Moscow region, Russia\\
% ...
$^{57}$ Southeast University, Nanjing 211100, People's Republic of China\\
$^{58}$ State Key Laboratory of Particle Detection and Electronics, Beijing 100049, Hefei 230026, People's Republic of China\\
% ...
$^{72}$ University of Science and Technology of China, Hefei 230026, People's Republic of China\\
% ...
$^{75}$ University of Turin and INFN, (A)University of Turin, I-10125, Turin, Italy; (B)University of Eastern Piedmont, I-15121, Alessandria, Italy; (C)INFN, I-10125, Turin, Italy\\
$^{76}$ Uppsala University, Box 516, SE-75120 Uppsala, Sweden\\
% ...
$^{81}$ Zhengzhou University, Zhengzhou 450001, People's Republic of China\\
\vspace{0.2cm}
$^{a}$ Deceased\\
% ...
$^{c}$ Also at the Novosibirsk State University, Novosibirsk, 630090, Russia\\
% ...
}
}

\begin{abstract}
By analyzing 7.93~$\rm fb^{-1}$ of $e^+e^-$ collision data collected at the center-of-mass energy of 3.773~GeV with the BESIII detector operated at the BEPCII collider, we report the observation of the semimuonic decays of $D^+\to \bar K_1(1270)^0\mu^+\nu_\mu$ and $D^0\to K_1(1270)^-\mu^+\nu_\mu$ with statistical significances of $12.5\sigma$ and $6.0\sigma$, respectively.
Their decay branching fractions are determined to be  ${\mathcal B}[D^{+}\to \bar{K}_1(1270)^0 \mu^{+}\nu_{\mu}]=(2.36\pm0.20^{+0.18}_{-0.27}\pm 0.48)\times10^{-3}$ and ${\mathcal B}[D^{0}\to K_1(1270)^{-} \mu^{+}\nu_{\mu}]=(0.78\pm0.11^{+0.05}_{-0.09}\pm 0.15)\times10^{-3}$, where the first and second uncertainties are statistical and systematic, respectively, and the third originates from the input branching
fraction of $\bar K_{1}(1270)^0\to K^- \pi^+\pi^0$ or $K_1(1270)^-\to K^-\pi^+\pi^-$.
Combining our branching fractions with the previous measurements of ${\mathcal B}[D^+\to \bar K_1(1270)^0e^+\nu_{e}]$ and ${\mathcal B}[D^0\to K_1(1270)^-e^+\nu_{e}]$, we determine the branching fraction ratios to be
${\mathcal B}[D^+\to \bar K_1(1270)^0\mu^+\nu_{\mu}]/{\mathcal B}[D^+\to \bar K_1(1270)^0e^+\nu_{e}]=1.03 \pm 0.14 \substack{+0.11\\-0.15}$ and
${\mathcal B}[D^0\to K_1(1270)^-\mu^+\nu_{\mu}]/{\mathcal B}[D^0\to K_1(1270)^-e^+\nu_{e}]=0.74\pm 0.13 \substack{+0.08\\-0.13}$.
Using the branching fractions  measured in this work and the world-average lifetimes of the $D^+$ and $D^0$ mesons, we determine the semimuonic partial decay width ratio to be $\Gamma [D^+\to \bar K_1(1270)^0 \mu^+\nu_\mu]/\Gamma [D^0\to K_1(1270)^- \mu^+\nu_\mu]=1.22\pm 0.10\substack{+0.06\\-0.09}$, which is consistent with unity as predicted by isospin conservation.
\end{abstract}

\maketitle

\oddsidemargin  -0.2cm
\evensidemargin -0.2cm

Experimental studies of the semileptonic (SL) decays of charmed mesons are important to deeply understand nonperturbative strong-interaction dynamics in weak decays.
The SL $D$ transitions into $\bar K_1(1270)$ offer an clean window to access the $K_1(1270)$ and $K_1(1400)$, which are mixtures of the $1^3P_1$ and $1^1P_1$ states of $K_1$ with mixing angle $\theta_{K_1}$. The determination of $\theta_{K_1}$ is essential for the theoretical description of $\tau$~\cite{Suzuki:1993yc}, $B$~\cite{Hatanaka:2008xj,Cheng:2004yj}, and $D$~\cite{Cheng:2010vk,Wang:2020pyy,Cheng:2003bn} decays into axial-vector strange mesons.
Throughout this paper, charged-conjugate modes are implied and $K_1$ denotes $K_1(1270)$ unless stated otherwise.

Branching fractions (BFs) for $D^{+(0)}\to \bar K_{1} \ell^+ \nu_\ell (\ell = e$ or $\mu)$ were calculated in the Isgur-Scora-Grinstein-Wise (ISGW) quark model~\cite{Isgur:1988gb} and its update with ISGW2~\cite{Scora:1995ty}.
The BFs for $D^{+}\to \bar K_{1} \mu^+ \nu_\mu$ and $D^{0}\to \bar K_{1} \mu^+ \nu_\mu$ were predicted to be 0.1\% and 0.3\%, respectively. However, this model ignores mixing between $^1\rm P_1$ and $^3\rm P_1$ states.
Recently, the BFs of $D^{+(0)}\to \bar K_{1} \ell^+ \nu_\ell$ were calculated with other theoretical approaches, including three-point quantum chromodynamics (QCD) sum rules (3PSR)~\cite{Khosravi:2008jw}, the covariant light-front quark model (CLFQM)~\cite{Cheng:2017pcq}, and light-cone QCD sum rules (LCSR)~\cite{Momeni:2019uag}.
In general, the predicted BFs range from $10^{-3}$ to $10^{-2}$~\cite{Khosravi:2008jw,Cheng:2017pcq,Momeni:2019uag,Bian:2021gwf}, and are sensitive to both the amplitude and sign of $\theta_{K_1}$.
Measurements of the BFs of $D^{+(0)}\to \bar K_1\mu^+\nu_\mu$ are important to test these theoretical calculations, to explore the nature of axial-vector strange mesons, and to understand the weak-decay mechanisms of $D$ mesons.

Lepton flavor universality (LFU) is a basic feature of the standard model (SM), in which the couplings between the three families of leptons and the gauge bosons are independent of lepton family. 
Nevertheless, a few hints of tension between experimental measurements and SM predictions have been reported in some cases, including SL $B$ decays~\cite{HFLAV} and the Cabibbo-angle anomaly~\cite{Coutinho:2019aiy,Crivellin:2020lzu}.
Intensive tests of LFU in different SL decays of heavy mesons are important to understand these anomalies.
Reference~\cite{Fajfer2015} notes that there may indeed be observable LFU violation effects in the SL decays mediated via $c\to s\ell^+\nu_\ell$.
In the SM, the ratio ${\mathcal R}^{\bar{K}_1}_{\mu/e}={\mathcal B}_{D\to \bar{K}_1 \mu^+\nu_\mu}/{\mathcal B}_{D\to \bar{K}_1 e^+\nu_e}$ is predicted to be $0.95-0.99$~\cite{Khosravi:2008jw,Cheng:2017pcq,Momeni:2019uag,Bian:2021gwf}.
To date, no experimental study of the semimuonic $D^{+(0)}$ decays into axial-vector mesons has been reported, and LFU in the SL $D^{+(0)}$ decays into axial-vector mesons has never been tested~\cite{Ke:2023qzc}.
Observation of the decays and measurements of the BFs of $D^{+}\to \bar{K}_1^0 \mu^{+}\nu_{\mu}$ and $D^{0}\to K_1^{-} \mu^{+}\nu_{\mu}$ offer an important opportunity to test $\mu$-$e$ LFU.

This paper reports the first observation of $D^+\to \bar K_1^0\mu^+\nu_\mu$ and $D^0\to K_1^-\mu^+\nu_\mu$ by using an $e^+e^-$ collision data sample corresponding to an integrated luminosity of 7.93~fb$^{-1}$~\cite{BESIII:2024luminosity},  recorded at the center-of-mass energy of $\sqrt s=3.773$ $\rm \,GeV$ with the BESIII detector~\cite{BESIII}.
Using our BFs and the world average BFs of $D^{+}\to \bar{K}_1^0 e^+\nu_e$, we provide the first test of the $\mu$-$e$ LFU with $D^{+}\to \bar{K}_1^0 \ell^+\nu_\ell$.
Combining the BFs of $D^+\to \bar K_1^0\mu^+\nu_\mu$ and $D^0\to K_1^-\mu^+\nu_\mu$ with the lifetimes of the $D^+$ and $D^0$, we test isospin invariance in $D\to \bar{K}_1 \mu^+\nu_\mu$ decays.

A description of the design and performance of the BESIII detector can be found in Ref.~\cite{BESIII}. The end cap TOF system was upgraded in 2015 using multigap resistive plate chamber technology, providing a time resolution of 60~ps~\cite{BESIII:TOF}, which benefits 67\% of the data used in this analysis.

Simulated data samples are produced with a {\sc geant4}-based~\cite{geant4} Monte Carlo (MC) toolkit including the geometric description of the BESIII detector and the detector response.
The simulation includes the beam energy spread and initial state radiation (ISR) in the $e^+e^-$ annihilations with the generator {\sc kkmc}~\cite{kkmc}.
In the MC simulation, the production of open-charm processes directly produced via $e^+e^-$ annihilations are modeled with the generator {\sc conexc}~\cite{conexc}.
The ISR production of vector charmonium(like) states and the continuum processes are incorporated in {\sc kkmc}~\cite{kkmc}.
All particle decays are modeled with {\sc evtgen}~\cite{evtgen} using BFs either taken from the Particle Data Group~\cite{pdg2022}, when available, or otherwise estimated with {\sc lundcharm}~\cite{lundcharm}.
Final state radiation from charged final state particles is incorporated using {\sc photos}~\cite{photos}. The signal MC events of $D^{+(0)}\to K^-\pi^+\pi^{0(-)}\mu^+\nu_{\mu}$ are generated with a special MC generator developed from the amplitude analysis of  $D^{+(0)}\to K^{-}\pi^{+}\pi^{0(-)}e^{+}\nu_{e}$~\cite{BESIII:k1ff}.

This work analyzes the $e^+e^-\to \psi(3770)\to D\bar{D}$ decay chain.
The single-tag (ST) candidate events are selected by reconstructing a  $D^-$ or $\bar D^0$ in the hadronic final states $D^- \to K^{+}\pi^{-}\pi^{-}$, $K^0_{S}\pi^{-}$, $K^{+}\pi^{-}\pi^{-}\pi^{0}$, $K^0_{S}\pi^{-}\pi^{0}$, $K^0_{S}\pi^{+}\pi^{-}\pi^{-}$, $K^{+}K^{-}\pi^{-}$, or $\bar D^0 \to K^+\pi^-$, $K^+\pi^-\pi^0$, $K^+\pi^-\pi^-\pi^+$.
These inclusively selected candidates are referred to as ST $D^-$ ($\bar D^0$) mesons. In the presence of the ST $D^-$ ($\bar D^0$) mesons, candidates for the signal decays are selected to form double-tag (DT) events.
The BF of the semimuonic decay is determined by
\begin{equation}
	\label{eq:bf}
	{\mathcal B}_{\rm SL}=N_{\mathrm{DT}}/(N_{\mathrm{ST}}^{\rm tot} \, \cdot\varepsilon_{\rm sig}\cdot\mathcal{B}_{K_{1}}),
\end{equation}
where $N_{\rm ST}^{\rm tot}$ and $N_{\rm DT}$ are the total ST and DT yields in the data sample, $\varepsilon_{\rm sig} = \Sigma_i [(\varepsilon^i_{\rm DT} \, N^i_{\rm ST})/(\varepsilon^i_{\rm ST} \, N^{\rm tot}_{\rm ST})]$ is the efficiency of detecting the semimuonic decay in the presence of the ST $D^-$ ($\bar D^0$) meson and $\mathcal{B}_{K_{1}}$ is the BF of $\bar K_1\to K^-\pi^+\pi^{0(-)}$.
Here, $i$ denotes the ST mode, and $\varepsilon_{\rm ST}$ and $\varepsilon_{\rm DT}$ are efficiencies for selecting the ST and DT candidates, respectively.

The selection criteria of $K^\pm$, $\pi^\pm$, $K^0_S$, and $\pi^0$ are the same as those used in Ref.~\cite{BESIII:st}.
The ST $\bar D$ ($\bar D$ denotes $D^-$ or $\bar D^0$) mesons are selected based on two kinematic variables in the $e^+e^-$ center-of-mass frame: the energy difference $\Delta E\equiv E_{\bar D}-E_{\mathrm{beam}}$ and the beam-constrained mass $M_{\rm BC}\equiv\sqrt{E_{\mathrm{beam}}^{2}/c^{4}-|\vec{p}_{\bar D}|^{2}/c^{2}}$.  Here, $E_{\mathrm{beam}}$ is the beam energy, and $E_{\bar D}$ and $\vec{p}_{\bar D}$ are the total energy and momentum of the $\bar D$ candidate, respectively. 
If there are multiple combinations for a given ST mode, the one giving the minimum $|\Delta E|$ is retained.

For each tag mode, the yield of reconstructed $\bar D$ mesons is obtained from an unbinned maximum likelihood fit to its $M_{\rm BC}$ distribution. In the fit, the signal distribution is modeled by the simulated shape convolved with a double-Gaussian resolution function to model the resolution difference between the data and the MC simulation. 
The combinatorial backgrounds are parametrized by an ARGUS function~\cite{ARGUS} with the endpoint fixed at $E_{\rm beam}=$1.8865~GeV.

The candidates with $M_{\rm BC}$ within (1.863,1.877) GeV/$c^2$ for ST $D^-$ and (1.859,1.873) GeV/$c^2$ for ST $\bar D^0$ are kept for further analyses.
For detailed information about the $\Delta E$ requirements, the ST yields in data, and the ST efficiencies, see Table 1 of the Supplemental Material~\cite{supplement}.
Summing over all tag modes, we obtain the total yields of ST $D^-$ and $\bar D^0$  mesons to be $(4149.9 \pm 2.3_{\rm stat})\times 10^3$ and $(6306.8 \pm 2.8_{\rm stat})\times 10^3$, respectively.

Candidates for $D^+\to \bar K_1^0\mu^+\nu_\mu$ or $D^0\to K_1^-\mu^+\nu_\mu$ are reconstructed from the remaining tracks and showers not used in the $D^-$ or $\bar{D}^0$ ST selection.  
The $\bar K_{1}^0$ or $K_1^-$ meson is reconstructed using its dominant decay $\bar K_{1}^0\to K^- \pi^+\pi^0$ or $K_1^-\to K^-\pi^+\pi^-$.
Candidates for $K^-$, $\pi^{\pm}$ and $\pi^0$ are selected with the same criteria as those used in the ST selection.
To identify muon candidates, we  use the information from ${\rm d}E/{\rm d}x$, the time-of-flight system~(TOF), and the electromagnetic calorimeter~(EMC), 
and combined confidence levels for the muon, pion, and kaon hypotheses ($CL_\mu$, $CL_{\pi}$ and $CL_{K}$) are calculated. 
Muon candidates are required to satisfy $CL_\mu>0.001$, $CL_\mu>CL_K$ and $CL_\mu>CL_\pi$.
To reduce the background from hadrons, muon candidates are also required to have a deposited energy in the EMC within less than 0.28 GeV.
The DT candidates must not contain any additional charged tracks, $N^{\rm char}_{\rm extra} = 0$; otherwise they are vetoed. 

To suppress the backgrounds containing extra $\pi^0$ mesons, such as the hadronic decays $D^+ \to K^- \pi^+ \pi^+ \pi^0 \pi^0$ and $D^0 \to K^- \pi^+ \pi^- \pi^+ \pi^0$, we require that there are no additional combinations of two photons that satisfy the requirements for a $\pi^0$ meson in the event selection, $N_{\rm extra}^{\pi^0} = 0$. The maximum energy of any extra photons ($E_{\text{extra~}\gamma}^{\rm max}$) that have not been used in reconstructing the signal final states is required to be less than 0.20 GeV. The opening angle between the missing momentum and the most energetic shower, $\theta_{\vec{p}_{\mathrm{miss}},\gamma}$, is required to satisfy $\cos\theta_{\vec{p}_{\mathrm{miss}},\gamma}<0.69\,(0.54)$ for $D^{+}\to K^-\pi^+\pi^0\mu^+\nu_{\mu}$ \, ($D^{0}\to K^-\pi^+\pi^-\mu^+\nu_{\mu}$).

The presence of the undetectable neutrino is inferred via the kinematic quantity $U_{\mathrm{miss}}\equiv E_{\mathrm{miss}}-|\vec{p}_{\mathrm{miss}}|c$, where $E_{\mathrm{miss}}$ and $\vec{p}_{\mathrm{miss}}$ are the missing energy and momentum of the semimuonic candidate, respectively, calculated by $E_{\mathrm{miss}}\equiv E_{\mathrm{beam}}-\Sigma_j E_j$ and $\vec{p}_{\mathrm{miss}}\equiv\vec{p}_{D}-\Sigma_j \vec{p}_j$ in the $e^+e^-$ center-of-mass frame. The index $j$ sums over the $K^-$, $\pi^+$, $\pi^{0(-)}$ and $\mu^+$ of the signal candidate, and $E_j$ and $\vec{p}_j$ are the energy and momentum of the $j$th particle, respectively.
In order to improve the $U_{\mathrm{miss}}$ resolution, the $D$ energy is constrained to the beam energy and $\vec{p}_{D} \equiv - \hat{p}_{\bar D}\sqrt{E_{\mathrm{beam}}^{2}/c^{2}-m_{D}^{2}c^{2}}$, where $\hat{p}_{\bar D}$ is the unit vector in the momentum direction of the ST $\bar D$, and $m_{D}$ is the nominal $D$ mass~\cite{pdg2022}.

In order to suppress backgrounds from the hadronic decays $D^{+(0)}\to K^-\pi^+\pi^+\pi^{0(-)}$, with misidentification of a pion as a muon, the invariant mass of the $K^-\pi^+\pi^{0(-)}\mu^+$ combination ($M_{K^-\pi^+\pi^{0(-)}\mu^+}$), is required to be less than 1.72\,(1.65)~GeV/$c^2$.
These selection criteria have been optimized by analyzing the inclusive MC samples with the figure-of-merit defined by $S/\sqrt{S+B}$, where $S$ and $B$ denote the signal and background yields, respectively.
	
\begin{figure}[htbp]
	\includegraphics[width=\linewidth]{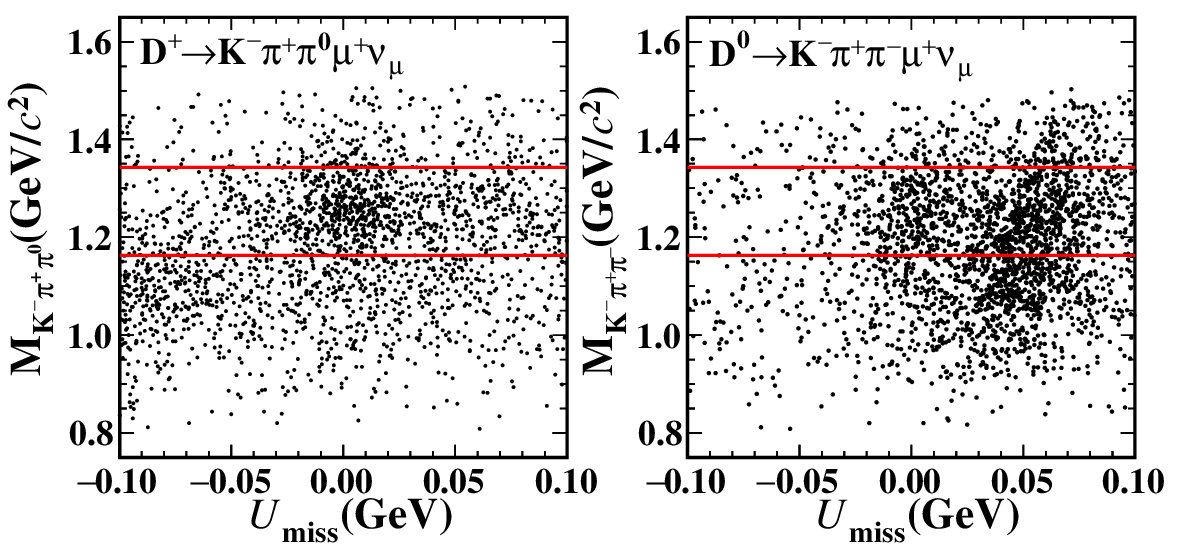}
	\caption{The $M_{K^-\pi^+\pi^{0(-)}}$ vs. $U_{\rm miss}$ distributions of the candidates for (left) $D^{+}\to K^-\pi^+\pi^0\mu^+\nu_{\mu}$ and (right) $D^{0}\to K^-\pi^+\pi^-\mu^+\nu_{\mu}$. The red lines denote the accepted mass range, $(1.163,1.343)$~GeV/$c^2$.}
	\label{fig:sig}
\end{figure}

\begin{figure}[htbp]
	\includegraphics[width=\linewidth]{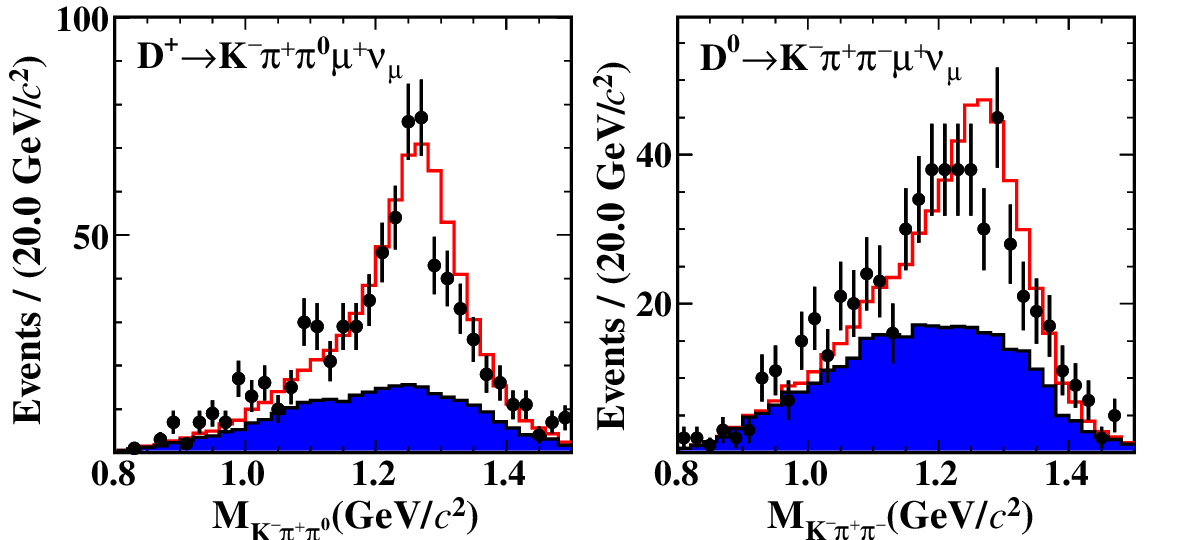}
	\caption{The $M_{K^-\pi^+\pi^{0}}$ (left) and $M_{K^-\pi^+\pi^{-}}$ (right) distributions of the candidates for $D^{+}\to K^-\pi^+\pi^0\mu^+\nu_{\mu}$ and $D^{0}\to K^-\pi^+\pi^-\mu^+\nu_{\mu}$. The points with error bars are data, the blue filled histograms are the simulated background, and the red line histograms are the signal MC samples. A requirement of $|U_{\rm miss}|<0.02$ GeV has been applied.}
	\label{fig:comm123}
\end{figure}

Figure~\ref{fig:sig} shows the distributions of $M_{K^-\pi^+\pi^{0(-)}}$ vs.~$U_{\rm miss}$ of the accepted candidates for $D^{+}\to K^-\pi^+\pi^0\mu^+\nu_{\mu}$ and $D^{0}\to K^-\pi^+\pi^-\mu^+\nu_{\mu}$ in data. 
Figure~\ref{fig:comm123} shows the distributions of $M_{K^-\pi^+\pi^{0(-)}}$ of $D^{+(0)}\to K^-\pi^+\pi^{0(-)}\mu^+\nu_\mu$  candidates, in which significant $K_1(1270)$ signals can be seen. 
Selected $K_1$ candidates must have $M_{K^-\pi^+\pi^{0(-)}}$ within the interval $(1.163,1.343)$~GeV/$c^2$, 
which corresponds to about $\pm 1 \Gamma$ (decay width) around the $K_1$ nominal mass~\cite{pdg2022}.

\begin{figure}[htbp]
	\includegraphics[width=\linewidth]{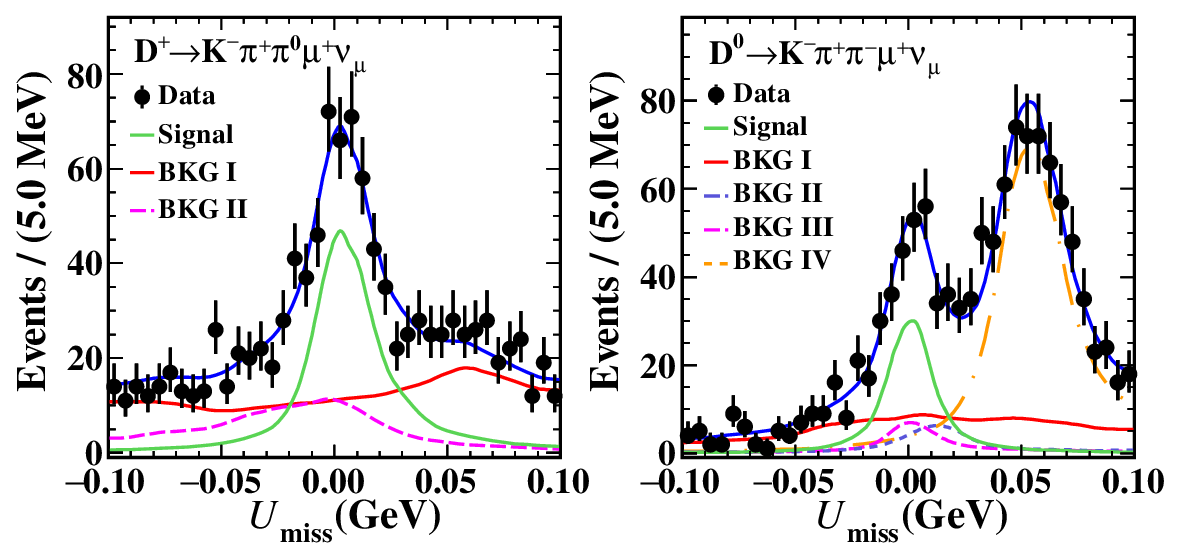}
	\caption{Fits to the $U_{\rm miss}$ distributions of the candidates for (left)  $D^{+}\to K^-\pi^+\pi^0\mu^+\nu_{\mu}$ and (right)  $D^{0}\to K^-\pi^+\pi^-\mu^+\nu_{\mu}$. The points with error bars are data. The blue solid curves are the total fits, the green solid curves denote the semimuonic signal shapes, and the red solid curves are from the combinatorial background shapes (BKG I). In the left plot, the magenta dashed curve is the background contribution from $D^{+}\to K^{-}\pi^{+}\pi^{+}\pi^{0}$ (BKG II). In the right plot, the violet and magenta dashed curves are the background contributions from $D^0\to K^-\pi^-\eta^{\prime}_{\pi^+\pi^-\gamma}$ (BKG II) and $D^0 \to K^-\pi^+K^+\pi^-$ (BKG III), respectively. The orange dashed curves is the background contribution from $D^0\to K^-\pi^+\pi^-\pi^+\pi^0$ (BKG IV).}
	\label{fig:fitsig}
\end{figure}

Figure~\ref{fig:fitsig} shows the distributions of $U_{\rm miss}$ of the accepted candidate events in data.
Clear signals around zero are observed.
The DT yields are extracted  from unbinned extended maximum-likelihood fits to the $U_{\rm miss}$ distributions of data. In the fits, the signal shapes are described by the MC-simulated shape extracted from the signal MC events and convolved with a Gaussian function due to the resolution difference between the data and MC simulation.  The combinatorial background shapes are modeled by the MC-simulated shape obtained from the inclusive MC samples.  The signal yield, the combinatorial background yield, and the Gaussian function's parameters are floated in the fits. 

For $D^{+}\to K^-\pi^+\pi^0\mu^+\nu_{\mu}$, the peaking background events from $D^{+}\to K^{-}\pi^{+}\pi^{+}\pi^{0}$ (BKG II) concentrate around 0 GeV;
for $D^{0}\to K^-\pi^+\pi^-\mu^+\nu_{\mu}$, the peaking background events from $D^0\to K^-\pi^-\eta^{\prime}_{\pi^+\pi^-\gamma}$ (BKG II),
$D^0 \to K^-\pi^+K^+\pi^-$ (BKG III), and $D^0\to K^-\pi^+\pi^-\pi^+\pi^0$ (BKG IV) concentrate around 0 GeV, 0 GeV, and 0.05 GeV, respectively.
All peaking background components are modeled by the MC-simulated shapes obtained from the corresponding MC samples. 
The backgrounds peaking near zero are convolved with the same Gaussian smearing function as the signal with their yields fixed based on MC simulations. The peaking around 0.05 GeV is convolved with a separate Gaussian function and its yield and parameters of Gaussian function are free.

In the fits, the $K^-\pi^+\pi^{0(-)}$ system is assumed to be dominated by the axial-vector meson $K_1$, while other components are ignored due to limited statistics (see Fig.~1, and 2 of supplemental materials~\cite{supplement}).
From the fits, we obtain the DT yields to be $382.8\pm31.9$ and $180.3\pm25.9$ for $D^{+}\to K^-\pi^+\pi^0\mu^+\nu_{\mu}$ and $D^{0}\to K^-\pi^+\pi^-\mu^+\nu_{\mu}$, respectively, where the uncertainties are statistical.
Their statistical significances are  estimated to be $14.0\sigma$ and $6.8\sigma$, respectively, by comparing the likelihoods with and without the signal components included, taking into account the change in the number of degrees of freedom. After further considering the non-$K_1$ component of $K^{-}\pi^{+}\pi^{0(-)}$, as discussed in the systematic uncertainty part, the significances of $D^{+}\to K^-\pi^+\pi^0\mu^+\nu_{\mu}$ and $D^{0}\to K^-\pi^+\pi^-\mu^+\nu_{\mu}$ are 12.5$\sigma$ and 6.0 $\sigma$, respectively.

The DT efficiencies, $\varepsilon^{i}_{\rm DT}$, obtained from signal MC samples with the same analysis procedures as for data, are  summarized in Table 1 of Ref.~\cite{supplement}. The averaged signal efficiencies in the presence of the ST $\bar D$ mesons are $(8.45\pm0.05)\%$ and $(10.92\pm0.07)\%$ for $D^{+}\to K^-\pi^+\pi^0\mu^+\nu_{\mu}$ and $D^{0}\to K^-\pi^+\pi^-\mu^+\nu_{\mu}$, respectively. These efficiencies include correction factors which are discussed later. The final signal efficiency, $\varepsilon_{\rm sig}$, includes the BF of $\pi^0\to\gamma\gamma$~\cite{pdg2022}.
The reliability of the MC simulation has been verified by the agreement of 
the distributions of the momenta and $\cos\theta$ of
$K^-$, $\pi^+$, $\pi^{0(-)}$, and $\mu^+$ with the data. 

By inserting $N_{\rm DT}$, $\varepsilon_{\rm sig}$, and $N_{\rm ST}^{\rm tot}$ into Eq.~(\ref{eq:bf}), we determine the product of ${\mathcal B}_{\rm SL}$ and ${\mathcal B}_{K_1}$~\cite{pdg2022} to be

\begin{equation*}
	{\mathcal B}^+_{\rm SL}{\mathcal B}^0_{K_1} = (10.92 \pm 0.91 \substack{+0.82\\-1.26})\times10^{-4},
\end{equation*}
\begin{equation*}
	{\mathcal B}^0_{\rm SL}{\mathcal B}^-_{K_1} = (2.62 \pm 0.38 \substack{+0.17\\-0.29})\times10^{-4},
\end{equation*}
where the first and second uncertainties are statistical and systematic, respectively. All related numbers are summarized in Table~\ref{table:bf}.

\begin{table*}[htbp]
	\centering
	\caption{The DT yields, $N_{\rm DT}$, in data, their statistical significances, the signal efficiencies, $\bar \epsilon_{\rm SL}$, and the obtained BFs, ${\mathcal B}_{\rm SL}$, for the two signal modes. The first and second uncertainties are statistical and systematic, respectively. The third uncertainty is from knowledge of the external BF, ${\mathcal B}_{K_1}$.}
	\label{table:bf}
	\resizebox{\linewidth}{!}{
	\begin{tabular}{lcccccc}
		\hline\hline
		Signal decay & $N_{\rm DT}$ & Significance ($\sigma$) & $\bar \varepsilon_{\rm SL}$ (\%) & ${\mathcal B}_{K_1} \, {\mathcal B}_{\rm SL}$ ($\times10^{-4}$) & ${\mathcal B}_{K_1}$ (\%) & ${\mathcal B}_{\rm SL}$ ($\times10^{-3}$)\\
		\hline
		$D^+\to \bar K_1^0\mu^+\nu_\mu$ & $382.8 \pm 31.9$ &$12.5$ & $8.45 \pm 0.05$ & $10.92\pm 0.91 \substack{+0.82\\-1.26}$ &$46.24 \pm 9.33$ & $2.36 \pm 0.20 \substack{+0.18\\-0.27} \pm 0.48$ \\
		$D^0\to K_1^-\mu^+\nu_\mu$      & $180.3 \pm 25.9$  &$6.0$ & $10.92\pm 0.07$ & $2.62 \pm 0.38 \substack{+0.17\\-0.29}$ &$33.74 \pm 6.54$ & $0.78 \pm 0.11 \substack{+0.05\\-0.09} \pm 0.15$ \\
		\hline\hline
	\end{tabular}
	}
\end{table*}

The systematic uncertainties in the BF measurement, which are assigned relative to the measured BFs, are discussed below.
The DT method ensures that most uncertainties arising from the ST selection cancel.
The uncertainty associated with the ST yield $N_{\rm ST}^{\rm tot}$, is assigned as 0.3\% after varying the signal and background shapes in the $M_{\rm BC}$ fit.

The uncertainties associated with the efficiencies of $\mu^+$ tracking, $K^-$ tracking and PID, $\pi^+$ tracking and PID, and $\pi^0$ reconstruction are investigated using data and MC samples of $e^+e^-\to\gamma \mu^+ \mu^-$ events and DT $D\bar D$ hadronic events ($\bar{D}^{0} \to K^{+} \pi^{-}$, $\bar{D}^{0} \to K^{+} \pi^{-} \pi^{0}$, and $\bar{D}^{0} \to K^{+} \pi^{+} \pi^{-} \pi^{-}$ vs. $D^{0}\to K^{-}\pi^{+}$, $D^{0}\to K^{-}\pi^{+}\pi^{+}\pi^{-}$, and $D^{0}\to K^{-}\pi^{+}\pi^{0}$, and $D^{-}\to K^{+}\pi^{-}\pi^{-}$ vs. $D^{+}\to K^{-}\pi^{+}\pi^{+}$).
We assign the uncertainties associated with the $\mu^+$ tracking, $K^-$ tracking (PID), $\pi^+$ tracking (PID) and $\pi^0$ reconstruction to be
1.0\%, 0.5\%\,(0.5\%), 0.5\%\,(0.5\%) and 2.0\%, respectively.

In the studies of $\mu^+$ PID efficiencies, the 2D (momentum and $\cos\theta$) PID efficiencies of data and MC simulation of $e^+e^-\to\gamma \mu^+ \mu^-$ events are reweighted to match those of $D^{+(0)} \to \bar K_1 \mu^+\nu_\mu$ decays.
The efficiency differences are $-(1.5 \pm 0.4)\%$ and $-(2.3\pm0.4)\%$ for $D^+\to \bar K_1^0\mu^+\nu_\mu$ and $D^0\to K_1^-\mu^+\nu_\mu$, respectively. The MC efficiencies are then corrected by these differences and used to determine the BFs of the semimuonic decays.
After corrections, we assign the uncertainties associated with the $\mu^+$ PID to be 0.4\% and 0.4\%, respectively.
The systematic uncertainties due to the combined requirements on $E_{\text{extra~}\gamma}^{\rm max}, N_{\rm extra}^{\pi^0}, N^{\rm char}_{\rm extra}$ and, separately, that on
$\cos\theta_{\vec{p}_{\mathrm{miss}},\gamma}$ are investigated using the control sample of $D^{+(0)}\to K_S^0\pi^{0(-)}\mu^+\nu_{\mu}$.
The corresponding systematic uncertainties are assigned as 3.9\% and 4.0\% for $D^+\to \bar K_1^0\mu^+\nu_\mu$ respectively; while those for $D^0\to K_1^-\mu^+\nu_\mu$ are 2.6\% and 2.3\%, respectively. 

To estimate the uncertainties due to the $M_{K^-\pi^+\pi^{0(-)}\mu^+}$ requirements, we use DT samples with
the signal modes replaced by  $D^+ \to K^0_S\pi^0\mu^+\nu_\mu$ and $D^0 \to K^0_S\pi^-\mu^+\nu_\mu$
and assign uncertainties of 4.2\% and 2.3\%, for the $D^+$ and $D^0$ modes, respectively.  

To estimate the uncertainties due to the $M_{K^-\pi^+\pi^{0(-)}}$ requirements, we remeasure the BFs with the nominal $K_1$ mass window half width of 0.90 GeV/$c^2$ widened to 1.10 GeV/$c^2$.  
The differences between the average BFs and the baseline values are assigned as the systematic uncertainties, which yields 0.6\% for both decay modes.  

The uncertainties associated with the fits to the $U_{\rm miss}$ distributions are estimated to be $\substack{+0.5\%\\-8.8\%}$ ($\substack{+4.4\%\\-10.0\%}$) which are dominated by the uncertainty from the background shape.
These systematic uncertainties are studied by altering the default MC background shape in three ways.
First, alternative MC samples are used to determine the background shape, where the relative fraction of the $q\bar q$ background is varied within the uncertainties~\cite{BESqqbar}.
Second, the BFs of the major $D\bar{D}$ background sources are varied by their uncertainties~\cite{pdg2022}.
Third, the non-$K_1^- (\bar{K}_1^0)$ sources of $K^-\pi^+\pi^- (K^-\pi^+\pi^0)$ ($\substack{+0.0\%\\-8.8\%}$ ($\substack{+0.0\%\\-9.0\%}$)) are the dominant uncertainties.
They are assigned to be the changes of the fitted DT yields after fixing nonresonant components by referring to the nonresonant fraction in $B\to J/\psi\bar{K}\pi\pi$~\cite{Belle:2010wrf}, since the nonresonant fraction in $D^{0(+)}\to\bar{K}\pi e^{+}\nu_{e}$~\cite{BESIII:2018jjm,BESIII:2015hty} is close to that in $B^+\to J/\psi\bar{K}\pi$~\cite{Belle:2002otd}.

The uncertainties related to the signal MC model are estimated to be 1.0\%\,(1.0\%), which correspond to the differences of the baseline signal efficiencies and those obtained with the signal MC samples after varying the input generator parameters by $\pm 1\sigma$.
The uncertainty due to the limited MC statistics is 0.4\%\,(0.4\%).

The total systematic uncertainties are 
$\substack{+7.5\%\\-11.6\%}$ and $\substack{+6.5\%\\-11.1\%}$ for $D^+\to \bar K_1^0\mu^+\nu_\mu$ and $D^0\to K_1^-\mu^+\nu_\mu$, respectively, obtained by adding all the individual contributions in quadrature. All systematic uncertainties are summarized in Table~\ref{table:sys}.

\begin{table}[htbp]
	\centering
	\caption{Systematic uncertainties in \% for the BF measurements.}
	\label{table:sys}
	\begin{tabular}{lcc}
\hline\hline
Source  & $\bar K_1^0\mu^+\nu_\mu$ & $K_1^-\mu^+\nu_\mu$  \\
\hline
$N^{\rm tot}_{\rm ST}$    & 0.3   & 0.3   	\\
$\mu^+$ tracking          & 1.0   & 1.0    	\\
$K^-, \pi^\pm$ tracking   & 1.0   & 1.5  	\\
$K^-, \pi^\pm$  PID       & 1.0   & 1.5    	\\
$\pi^0$ reconstruction    & 2.0   & ...    	\\
$\mu^+$  PID         	  & 0.4   & 0.4       	\\
$E_{\rm extra\gamma}^{\rm max}, N_{\pi^0}^{\rm extra}, N^{\rm extra}_{\rm char}$                         & 3.9   & 2.6         \\
$\cos\theta_{\vec{p}_{\mathrm{miss}},\gamma}$
                          & 4.0   & 2.3   	\\
$M_{K^-\pi^+\pi^{0(-)}\mu^{+}}$ requirement
                          & 4.2	  & 2.3  	\\
$M_{K^-\pi^+\pi^{0(-)}}$ requirement
                          & 0.6   & 0.6   	\\
$U_{\rm miss}$ fit        & $\substack{+0.5\\-8.8}$    & $\substack{+4.4\\-10.0}$  				\\
Signal model              & 1.0   & 1.0 	\\
MC statistics             & 0.4   & 0.4   	\\
\hline
Total                     & $\substack{+7.5\\-11.6}$   	& $\substack{+6.5\\-11.1}$  			\\
		\hline\hline
	\end{tabular}
\end{table}

Using the world average of ${\mathcal B}_{K_1\to K^-\pi^+\pi^0} = (46.24 \pm 9.33)\%$~\cite{K10}\,${\mathcal B}_{K_1\to K^-\pi^+\pi^-} = (33.74 \pm 6.54)\%$~\cite{K1m}, we obtain:
\begin{equation*}
	\small
	{\mathcal B}_{D^+\to\bar K_1^0\mu^+\nu_\mu} = (2.36\pm 0.20\substack{+0.18\\-0.27}\pm 0.48)\times10^{-3},
\end{equation*}
\begin{equation*}
	{\mathcal B}_{D^0\to K_1^-\mu^+\nu_\mu} = (0.78\pm 0.11\substack{+0.05\\-0.09}\pm 0.15)\times10^{-3},
\end{equation*}
where the third uncertainty is from the input BFs.

To summarize, by analyzing 7.93~$\rm fb^{-1}$ of $e^+e^-$ collisions data taken at $\sqrt s=3.773$~GeV,
we report the first observation of $D^+\to \bar K_1^0\mu^+\nu_\mu$ and $D^0\to K_1^-\mu^+\nu_\mu$ with statistical significances of $12.5\sigma$ and $6.0\sigma$, respectively. Their decay BFs are determined to be 
${\mathcal B}[D^{+}\to \bar{K}_1^0 \mu^{+}\nu_{\mu}]=(2.36\pm 0.20\substack{+0.18\\-0.27}\pm 0.48)\times10^{-3}$ and ${\mathcal B}[D^{0}\to K_1^{-} \mu^{+}\nu_{\mu}]=(0.78\pm 0.11\substack{+0.05\\-0.09}\pm 0.15)\times10^{-3}$.
Table~\ref{table:bfexp} shows the comparison of the measured BFs with several theoretical calculations. 
Because the BFs predicted in theory are sensitive to the $K_1$ mixing angle, our BFs provides some information on the $K_1$ mixing angle.  Our BFs disfavor the theoretical calculations from LCSR~\cite{Momeni:2019uag} and CLFQM~\cite{Bian:2021gwf}, which are under the assumption of the mixing angle $\theta_{K_1} = -(34 \pm 13)^{\circ}$, by more than $5\sigma$.

\begin{table*}[htbp]
	\centering
	\caption{Comparison of the BFs measured  in this work with the predicted values for $D^{+(0)}\to  \bar K_1\mu^+\nu_{\mu}$ ($\times 10^{-3}$) from various theories with differences given in units of standard deviations.}
	\label{table:bfexp}
	\resizebox{\linewidth}{!}{
		\begin{tabular}{llccccccccc}
			\hline\hline
			&                      & CLFQM~\cite{Cheng:2017pcq} & 3PSR~\cite{Khosravi:2008jw} & LCSR~\cite{Momeni:2019uag} & LCSR~\cite{Momeni:2019uag} & LCSR~\cite{Momeni:2019uag} & CLFQM~\cite{Bian:2021gwf} & LFQM~\cite{Bian:2021gwf}  & LCSR~\cite{Bian:2021gwf} &  This work                                                      \\ \hline
			Decay                                                    & $\theta_{K1}$        & $33^{\circ}$               & $37^{\circ}$                 & \multicolumn{3}{c}{$-(34\pm13)^{\circ}$}    &\multicolumn{3}{c}{$45^{\circ}$}                                         & --                                                            \\ \hline
			\multirow{2}{*}{$D^+\to \bar K^0_1\mu^+\nu_{\mu}$} & Predicted            & $2.60\pm0.30$              & $2.70\pm0.25$                & $18.59\pm0.31$             & $16.86\pm0.27$             & $19.73\pm0.32$     &  $5.9\substack{+0.47\\-0.11}$   &   $3.7\substack{+0.00\\-0.41}$   &   $1.4\substack{+0.15\\-0.27}$      & \multirow{2}{*}{$2.36\pm 0.20\substack{+0.18\\-0.27}\pm 0.48$} \\
			& Deviation ($\sigma$) & 0.4                        & 0.6                          & 27.3                      & 25.2                       & 29.0                       & 6.8                        & 2.6                & 1.8                             &                                                          \\ \hline
			\multirow{2}{*}{$D^0\to  {K^-_1}\mu^+\nu_{\mu}$}   & Predicted            & --                         & $1.03\pm0.1$                 & $8.09\pm0.15$              & $6.78\pm0.12$              & $8.92\pm0.16$      &  $2.3\substack{+0.19\\-0.44}$   &   $1.5\substack{+0.00\\-0.16}$   &   $0.54\substack{+0.06\\-0.11}$     & \multirow{2}{*}{$0.78\pm 0.11\substack{+0.05\\-0.09}\pm 0.15$} \\
			& Deviation ($\sigma$) &  --                        & 1.1                          & 28.6                      & 25.1                       & 31.1                       & 5.4                        & 2.8                & 1.1                             &                                                          \\ \hline\hline
		\end{tabular}
	}
\end{table*}

Combining the BFs measured in this work and the BFs of ${\mathcal B}[D^+\to\bar K_1^0e^{+}\nu_{e}] = (2.30\pm{0.26\substack{+0.18\\-0.21}} \pm 0.25) \times10^{-3}$~\cite{BESIII:2019dptok1} and ${\mathcal B}[D^0\to K_1^-e^{+}\nu_{e}] = (1.09\pm{0.13\substack{+0.09\\-0.13}} \pm 0.12) \times10^{-3}$~\cite{BESIII:2021dotok1} reported in the previous BESIII analysis, we determine the BF ratios ($\mathcal R^{D}_{\mu/e} = \frac{{\mathcal B}[D\to\bar K_1\mu^+\nu_{\mu}]}{{\mathcal B}[D\to\bar K_1e^+\nu_{e}]}$) to be
$\mathcal R^{D^+}_{\mu/e} = 1.03 \pm 0.14 \substack{+0.11\\-0.15}$ and $\mathcal R^{D^0}_{\mu/e} = 0.74\pm 0.13 \substack{+0.08\\-0.13}$. Here, the systematic uncertainties in ST yields, $K^{-}/\pi^{\pm}$ tracking and PID, and $\pi^{0}$ reconstruction cancel.
These are consistent with theoretical predictions based on $\mu$-$e$ LFU in $D^{+(0)}\to \bar K_1\ell^+\nu_\ell$ decays within current  uncertainties.
Using the BFs of $D^0\to K_1^-\mu^+\nu_\mu$ and $D^+\to \bar K_1^0\mu^+\nu_\mu$ measured in this work as well as the world-average lifetimes of the $D^0$ and $D^+$ mesons~\cite{pdg2022}, we determine the partial decay width ratio $\Gamma [D^+\to \bar K_1^0 \mu^+\nu_\mu]/\Gamma [D^0\to K_1^- \mu^+\nu_\mu]=1.22\pm 0.10\substack{+0.06\\-0.09}$,
which is consistent with unity as predicted by isospin conservation.

\section{Acknowledgements}
The BESIII Collaboration thanks the staff of BEPCII and the IHEP computing center for their strong support. This work is supported in part by National Key R\&D Program of China under Contracts Nos. 2023YFA1606000; National Natural Science Foundation of China (NSFC) under Contracts Nos. 11635010, 11735014, 11935015, 11935016, 11935018, 12025502, 12035009, 12035013, 12061131003, 12192260, 12192261, 12192262, 12192263, 12192264, 12192265, 12221005, 12225509, 12235017, 12361141819; the Chinese Academy of Sciences (CAS) Large-Scale Scientific Facility Program; the CAS Center for Excellence in Particle Physics (CCEPP); Joint Large-Scale Scientific Facility Funds of the NSFC and CAS under Contract No. U1832207; 100 Talents Program of CAS; The Institute of Nuclear and Particle Physics (INPAC) and Shanghai Key Laboratory for Particle Physics and Cosmology; German Research Foundation DFG under Contracts Nos. 455635585, FOR5327, GRK 2149; Istituto Nazionale di Fisica Nucleare, Italy; Ministry of Development of Turkey under Contract No. DPT2006K-120470; National Research Foundation of Korea under Contract No. NRF-2022R1A2C1092335; National Science and Technology fund of Mongolia; National Science Research and Innovation Fund (NSRF) via the Program Management Unit for Human Resources \& Institutional Development, Research and Innovation of Thailand under Contract No. B16F640076; Polish National Science Centre under Contract No. 2019/35/O/ST2/02907; The Swedish Research Council; U. S. Department of Energy under Contract No. DE-FG02-05ER41374.

\onecolumngrid
\appendix
\section*{Supplemental materials}
\label{supplemental}
Table~\ref{tab:st} summarizes some analysis details for the single-tag $D^-$ candidates, namely the $\Delta E$ requirements, the ST yields in data, the ST efficiencies, and the double-tag efficiencies. 

\begin{table}[htbp]
	\centering
	\caption {\label{tab:st}
		The $\Delta E$ requirements, the obtained ST $\bar D^0(D^-)$ yields ($N^{i}_{\rm ST}$) in data, the ST efficiencies ($\varepsilon^{i}_{\rm ST}$), and the DT efficiencies ($\varepsilon^{i}_{\rm DT}$). The efficiencies do not include the branching fractions of the $K^0_S$ and $\pi^0$ decays. The uncertainties are statistical only.}
	\begin{tabular}{lcccc}
		\hline\hline
		Tag mode                                      & $\Delta E$~(GeV)   &  $N^{i}_{\rm ST}~(\times10^{3})$         &  $\varepsilon^{i}_{\rm ST}~(\%)$   &  $\varepsilon^{i}_{\rm 
			DT}~(\%)$ \\\hline
		$D^{-} \to K^{+} \pi^{-} \pi^{-}$                 &$(-0.025,0.024)$   &   $2164.1\pm1.6$   &   $51.17\pm0.01$   &   $8.87\pm0.09$ \\
		$D^{-} \to K_{S}^{0} \pi^{-}$                     &$(-0.025,0.026)$   &   $250.4\pm0.5$   &   $50.74\pm0.02$   &   $9.17\pm0.09$ \\
		$D^{-} \to K^{+} \pi^{-} \pi^{-} \pi^{0}$         &$(-0.057,0.046)$   &   $689.0\pm1.2$   &   $25.50\pm0.01$   &   $7.74\pm0.12$ \\
		$D^{-} \to K_{S}^{0} \pi^{-} \pi^{0}$             &$(-0.062,0.049)$   &   $558.5\pm0.9$   &   $26.28\pm0.01$   &   $8.47\pm0.12$ \\
		$D^{-} \to K_{S}^{0} \pi^{-} \pi^{-} \pi^{+}$     &$(-0.028,0.027)$   &   $300.5\pm0.7$   &   $29.01\pm0.01$   &   $8.13\pm0.12$ \\
		$D^{-} \to K^{+} K^{-} \pi^{-}$                   &$(-0.024,0.023)$   &   $187.4\pm0.5$   &   $41.06\pm0.02$   &   $7.83\pm0.10$ \\
		\hline
		$\bar{D}^{0} \to K^{+} \pi^{-}$                   &$(-0.027,0.027)$   &   $1449.3\pm1.3$   &   $65.34\pm0.01$   &   $11.72\pm0.09$ \\
		$\bar{D}^{0} \to K^{+} \pi^{-} \pi^{0}$           &$(-0.062,0.049)$   &   $2913.2\pm2.0$   &   $35.59\pm0.01$   &   $11.16\pm0.12$ \\
		$\bar{D}^{0} \to K^{+} \pi^{+} \pi^{-} \pi^{-}$   &$(-0.026,0.024)$   &   $1944.2\pm1.6$   &   $40.83\pm0.01$   &   $9.95\pm0.11$ \\
		\hline\hline
	\end{tabular}
\end{table}

\begin{figure}[htbp]
	\includegraphics[width=\linewidth]{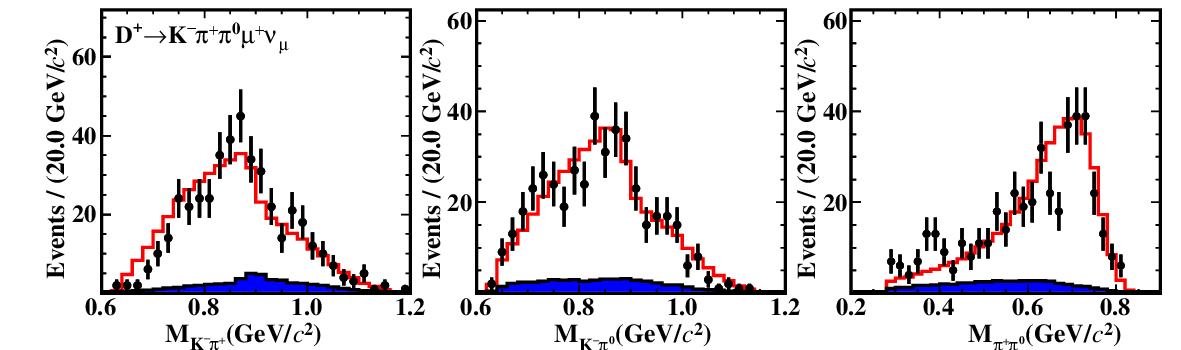}
	\caption{Comparison of $M_{K^-\pi^+}$ (left), $M_{K^-\pi^0}$ (center) and $M_{\pi^{+}\pi^{0}}$ (right) of the candidates for $D^{+}\to K^-\pi^+\pi^0\mu^+\nu_{\mu}$. The points with error bars are data, the blue filled histograms are the simulated background, and the red line histograms are the signal MC samples. A requirement of $|U_{\rm miss}|<0.02$ GeV has been applied.}
	\label{fig:commdp}
\end{figure}

\begin{figure}[htbp]
	\includegraphics[width=\linewidth]{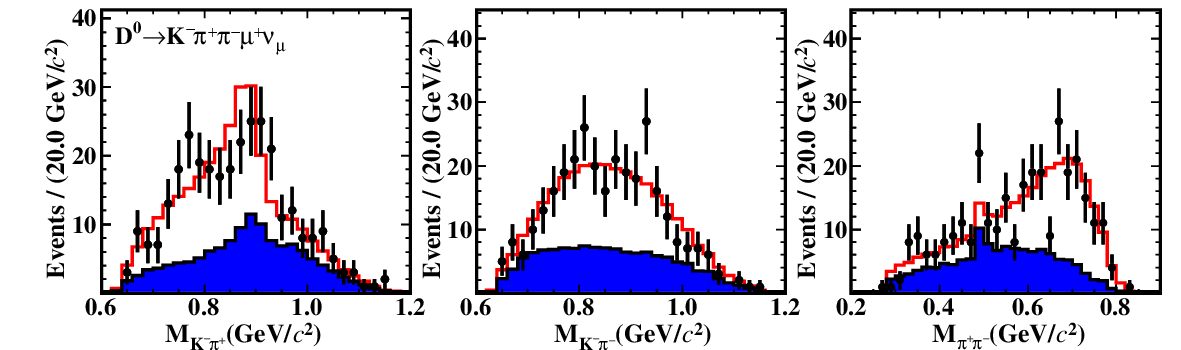}
	\caption{Comparison of $M_{K^-\pi^+}$ (left), $M_{K^-\pi^-}$ (center) and $M_{\pi^{+}\pi^{-}}$ (right) of the candidates for $D^{0}\to K^-\pi^+\pi^-\mu^+\nu_{\mu}$. The points with error bars are data, the blue filled histograms are the simulated background, and the red line histograms are the signal MC samples. A requirement of $|U_{\rm miss}|<0.02$ GeV has been applied.}
	\label{fig:commd0}
\end{figure}

Figure~\ref{fig:commdp} and~\ref{fig:commd0} shows the distributions of $M_{K^-\pi^+}$, $M_{K^-\pi^{0(-)}}$
and $M_{\pi^+\pi^{0(-)}}$ for $D^{+}\to K^-\pi^+\pi^0\mu^+\nu_{\mu}$ and $D^{0}\to K^-\pi^+\pi^-\mu^+\nu_{\mu}$ candidates, respectively.  Results from the data and the inclusive MC sample are overlaid.  

\end{document}